# Characterization of Optical Frequency Transfer Over 154 km of Aerial Fiber

DAVID R. GOZZARD,[1,2] SASCHA W. SCHEDIWY,[2,1,*] BRUCE WALLACE,[3] ROMEO GAMATHAM[4] AND
KEITH GRAINGE[5]

[1]School of Physics and Astrophysics, University of Western Australia, Perth, WA 6009, Australia
[2]International Centre for Radio Astronomy Research, University of Western Australia, Perth, WA 6009, Australia
[3]SKA South Africa, 17 Baker Street, Rosebank, Johannesburg, South Africa
[4]SKA South Africa, 3rd Floor, The Park, Park Road, Pinelands 7405, South Africa
[5]Jodrell Bank Centre for Astrophysics, Alan Turing Building, School of Physics & Astronomy, The University of Manchester,
Oxford Road, Manchester M13 9PL, UK
*Corresponding author: sascha.schediwy@uwa.edu.au

**We present measurements of the frequency transfer stability and analysis of the noise characteristics of an optical signal propagating over aerial suspended fiber links up to 153.6 km in length. The measured frequency transfer stability over these links is on the order of $10^{-11}$ at an integration time of one second dropping to $10^{-12}$ for integration times longer than 100 s. We show that wind-loading of the cable spans is the dominant source of short-timescale noise on the fiber links. We also report an attempt to stabilize the optical frequency transfer over these aerial links.**

http://dx.doi.org/10.1364/OL.99.099999

Long-distance optical fiber networks are increasingly being used for the transmission of high-precision frequency and timing signals because the stability and accuracy of optical links surpasses that of conventional satellite two-way time and frequency transfer by more than three orders of magnitude [1-3]. The frequency and timing signals have applications in science and industry ranging from radio astronomy, geodesy, and tests of fundamental physics, to network synchronization and high-precision manufacturing [4].

One factor that limits the performance of long-distance frequency transfer is the phase-noise imposed onto the optical signal by mechanical stresses on the fiber link. Measurements of frequency transfer stability have mostly focused on underground [5-7] and some submarine links [8]. Although aerial suspended fibers and optical ground wires (OPGWs) are subject to greater mechanical stresses due to their greater exposure, the lower cost of reticulating overhead fiber means that there are circumstances under which the timing and frequency transfer stability of transmissions employing overhead fiber links is of interest. In particular, this study was carried out in order to characterize the impact of synchronization signals that are to be disseminated over aerial fiber links as part of the Square Kilometre Array (SKA) radio telescope to be constructed in the Karoo region of South Africa [9].

Investigation of detrimental effects on telecommunications and network monitoring due to the impact of environmental conditions on aerial fibers and OPGWs has focused on polarization mode dispersion, state of polarization fluctuations, and transmission delay variations [10-14]. To the best of our knowledge, only one study [15] has undertaken a characterization of frequency transfer stability on aerial fiber to date, and only at radio frequencies (10 MHz). In this paper, we present what we believe to be the first characterization of optical frequency transfer stabilities on aerial suspended fiber links of 32.6 km, 65.2 km, and 153.6 km lengths.

Optical frequency transfer provides the highest level of precision for the purposes of metrology and other sciences [16], however, it is also the most sensitive to mechanical stresses on the link and is the most difficult to actively stabilize. The magnitude of the phase perturbations detected on the aerial fiber links in this study demonstrate the challenges for optical phase stabilization systems attempting to compensate for the noise imposed on the link.

These tests were conducted at SKA South Africa's Klerefontein support base near the South African SKA site. Figure 1 shows the physical arrangement of the relevant locations. Four cores of a 16.3 km fiber link from Klerefontein to the Carnarvon point-of-presence (POP) site, and two cores of a 76.8 km fiber link between Klerefontein and the SKA central site were used for the tests (Fiber standard: SMF E9 according to ITU-T G.652.D, chromatic dispersion at 1550 nm: 18 ps/nm.km). The fiber cables were suspended from power poles for the entirety of their runs. Fiber patch leads were installed at the Carnarvon POP to give two "loop-back" sections of 32.6 km of aerial fiber, while another patch, incorporating an *IDIL Fibres Optiques* bi-directional optical amplifier (with a gain of +18.7 dB and a noise figure of 7.6 dB), was installed at the SKA central site to produce a loop-back length of 153.6 km.

Figure 2 shows a schematic representation of the system that was installed at Klerefontein and used to characterize the fiber links.

The system is fundamentally a stabilized optical frequency transmission system, based on the technology pioneered by Ma et al. [17]. Our primary aim was to characterize the noise on the fiber, but we designed the equipment to also be capable of active stabilization of the link noise. However, in practice, we were unable to achieve continuously stabilized transmission for more than several minutes, even over the 32.6 km link, due to the magnitude of the perturbations and the design of the stabilization servo electronics.

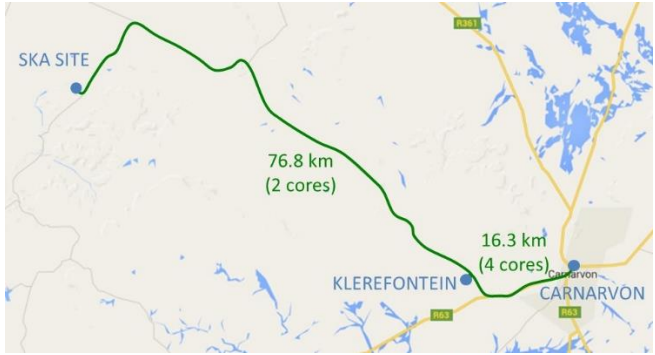

Fig. 1. Relevant locations and overhead fiber link routes investigated in this experiment. (Modified from Google Maps image.)

In the transmitter section of the system, an NKT Photonics *Koheras BASIK X15* commercial diode laser (spectral linewidth < 100 Hz) was used to provide a highly coherent optical frequency at 193 THz (1552 nm). A small fraction (3%) of the optical signal was split off to provide the out-of-loop optical reference for the measurement photodetector (PD). The rest of the power continued down-line where it was split again, with part being reflected off a Faraday mirror (FM) at the end of a short length of fiber to become the reference signal for an imbalanced Michelson interferometer. The fiber link constituted the long arm of the Michelson interferometer. Before the optical signal was injected into the fiber link, it passed through the servo acousto-optic modulator (AOM) which produced a +70 MHz shift of the optical frequency.

The transmitted signal made the round-trip through the fiber link and entered the receiver side of the system. The optical signal passed through a polarization controller (Pol) and was split, part of the power being directed to the measurement PD where it beat against the optical reference to produce a 70 MHz electronic signal, the stability of which, and thus the noise on the fiber link, was measured using a frequency counter (*Agilent 53132A*) which produced a triangle-weighted estimate of the fractional frequency stability. The polarization controller was used to align the polarization of the received signal with the polarization of the optical reference.

The rest of the received signal power passed through the receiver AOM (+50 MHz frequency shift), reflected from a FM, passed through the receiver AOM again and returned through the fiber link back to the transmitter side of the system. After passing through the servo AOM, the signal from the link and the signal from the reference arm, entered the input of the servo PD where they formed a 240 MHz electronic beat signal. This beat signal encoded information about the frequency fluctuations on the fiber link due to environmental perturbations. The electronic signal was divided by a factor of 24 and then mixed with a 10 MHz local oscillator. The mixer product became the servo error signal and was used to steer the 70 MHz output of the frequency synthesizer. Activating the frequency modulation closed the servo loop and the system adjusted the frequency shift produced by the servo AOM to compensate for frequency fluctuations caused by perturbations on the fiber link. Characterization of the noise of the fiber link was achieved with the servo loop deactivated. Links of 32.6 km (2×16.3 km, 8.6 dB loss) and 65.2 km (4×16.3 km, 17.4 dB loss) between Klerefontein and Carnarvon, and 153.6 km (2×76.8 km, 31.7 dB loss) between Klerefontein and the SKA site were tested during these trials.

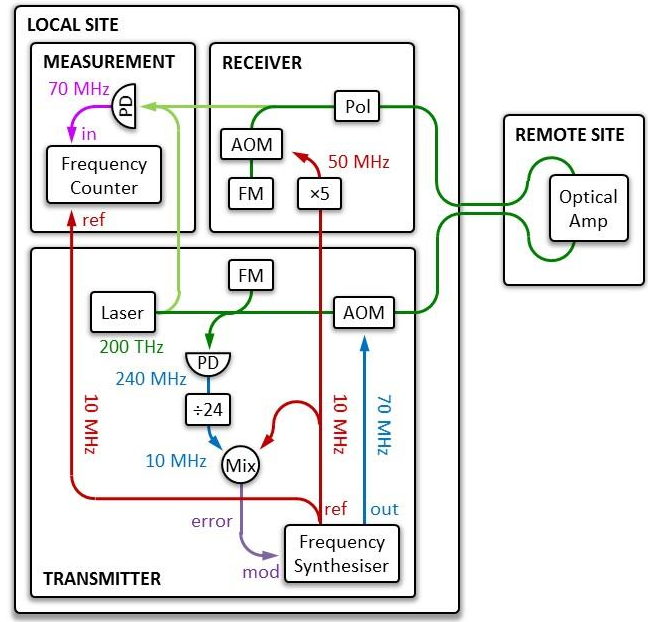

Fig. 2. Schematic of the fiber characterization system and 2×76.8 km fiber link with optical amplifier. PM, power meter; PD, photodetector; Pol, polarization controller; AOM, acousto-optic modulator; FM, Faraday mirror; Mix; frequency mixer; Ref, 10 MHz frequency reference; mod, frequency modulation input.

Weather data, including wind speed and wind direction, coinciding with the period of the trials were collected from a weather station operated at Klerefontein by the C-Band All Sky Survey (C-BASS). The wind velocity data were compared with the magnitude of the coincident link frequency fluctuations.

The data from the frequency counter were processed to produce fractional frequency stability curves for the three links studied. These curves are shown in Fig. 3. The blue trace in Fig. 3 (filled triangles, solid line) is a "zero-length" (a 2 m fiber patch lead) measurement with optical attenuation set equal to that of the 32.6 km link and is, therefore, a measurement of the fiber characterization system noise floor.

Due to the magnitude of the noise on the fiber links, it was not possible to actively stabilize the optical transmission for more than a few minutes before a cycle-slip occurred. The data presented here shows only one stabilized transmission of 26 minutes for the 32.6 km link (green, filled triangles, solid line). The measurements for the characterization of the frequency noise on the free-running links were taken for periods of between 1 and 93 hours at all times of day and night.

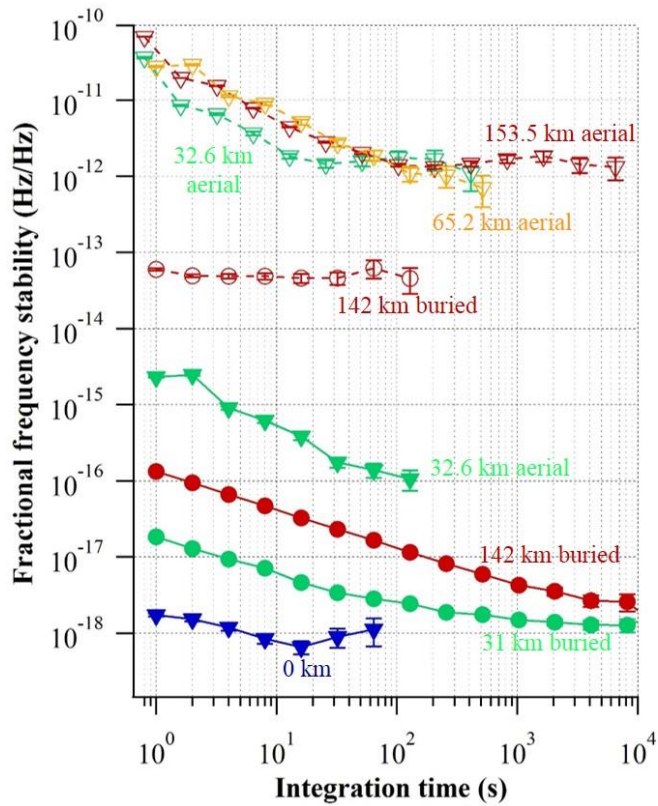

Fig. 3. Fractional frequency stability for the links under test derived from the Agilent 53132A frequency counter. Open marker with dashed lines — unstabilized transfer, closed marker with solid lines — stabilized transfer.

There is little difference in the overall frequency noise levels between the three different aerial link lengths (Fig. 3). However, all three links show an extreme level of noise. The instability of the 153.6 km link (red, open triangles, dashed line) at one second integration time is approximately 260 times greater than the free-running link noise for the 1840 km buried fiber link reported by Droste and colleagues [18].

For comparison with the stabilized 32.6 km overhead fiber data (green, filled triangles, Fig. 3), the stabilized transfer stability over a 31 km urban trench and conduit fiber link in Perth, Western Australia has been included (green, filled circles, solid line). Figure 3 shows that the fractional frequency value for the stabilized transfer over the 32.6 km aerial fiber link is 124 times greater than for the 31 km buried link at an integration time of one second, and that the value of the free-running aerial link (green, open triangles, dashed line) is nearly two million times greater than the stabilized buried link at an integration time of one second.

Also shown in Fig. 3 is the stability of a 142 km link, stabilized (red, filled circles, solid line) and free-running (red, open circles, dashed line), comprising 31 km of urban trench and conduit fiber, and 111 km of fiber spool in the laboratory in Perth. Compared to the 153.6 km Klerefontein-SKA site loop-back link (8% longer), the fractional frequency stability of the aerial link is nearly three orders of magnitude greater than the 142 km free-running link at an integration time of one second.

The extreme difference in the frequency transfer stability between aerial and buried fiber links is due to mechanical stresses on the fiber cables caused by the weather. During the testing period, the fiber cables were subjected to high- and gusting winds, rain, and large thermal gradients as the fiber went from overnight frost to direct morning sunlight. The greatest source of short-timescale perturbations was the wind.

The raw data from the frequency counter displayed a noticeable periodicity in the amplitude of the frequency fluctuations. By taking the Fourier transform, shown in Fig. 4, of the time-series frequency data for the 153.6 km Klerefontein-SKA site link, it can clearly be seen that there is a dominant periodicity in the frequency perturbations of 1.405±0.002 s (2σ).

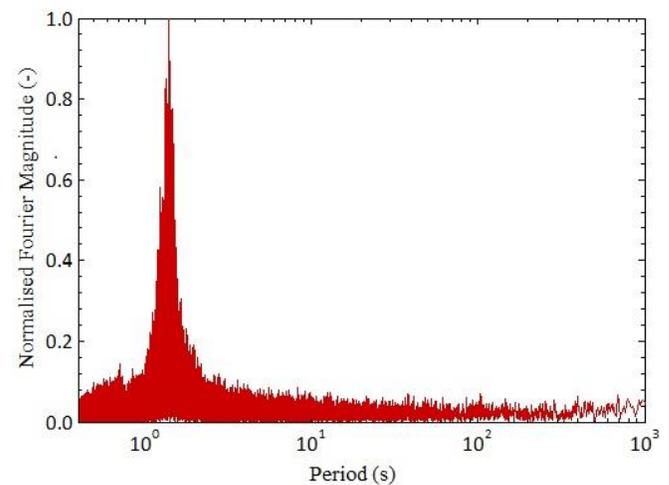

Fig. 4. Fourier transform of frequency over time data, showing a dominant 1.405 s period term.

Visual inspection of typical spans of aerial fiber in the vicinity of Klerefontein (recorded on camera for later analysis) estimated the period of the fundamental lateral (swinging) mode of the spans to be 2.85±0.15 s, giving a half-period of 1.43±0.08 s. This overlaps, within uncertainty, with the dominant period in the frequency fluctuation data (Fig. 4), and indicates that the swinging action of the cable is related to the periodicity of the frequency fluctuations. The swinging of the spans induces periodic stresses on the fiber, altering the optical path length and producing phase noise on the signal.

The Klerefontein-SKA aerial fiber link comprises around 1000 span-segments with a median span distance of 70 m. Because the frequency fluctuations induced by the mechanical oscillations of the fiber manifest as twice the cable fundamental mode frequency their signal magnitudes combine in quadrature, even when the oscillations of separate fiber cables spans are not coherent.

The impact of wind-loading on the fiber cables can be further examined by assessing the correlation between the magnitude of the frequency fluctuations (which are dominated by the 1.405 s periodic component) and the amplitude of the swinging of the cable spans. This oscillation mode is loaded by the vector component of the wind perpendicular to the direction of the cable span. Figure 5 shows a plot of the absolute magnitude of the frequency fluctuations over time, overlayed with the projected component of the wind speed (determined from C-BASS weather station data) perpendicular to the weighted dominant direction of the Klerefontein-SKA site fiber link.

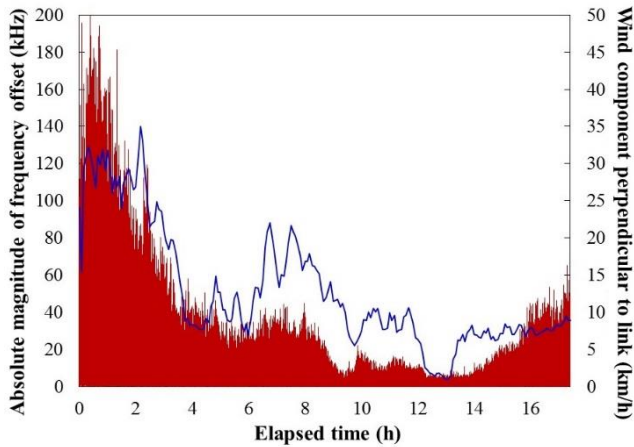

Fig. 5. Absolute magnitude of frequency fluctuations over time (filled red area), overlayed with the projected component of the wind speed perpendicular to the weighted dominant direction of the Klerefontein-SKA site fiber link (blue line).

The data suggests that there is a causal correlation between the wind speed and the magnitude of the frequency fluctuations. The residual discrepancy is attributed to the local variation in wind speed between the weather station site and the fiber route. The time resolution of the wind speed data also mask the effects of gusts.

We have measured the frequency transfer stability and characterized the noise of optical signals transmitted over aerial fiber links up to 153.6 km in length. All of the aerial fiber links tested exhibited levels of frequency noise hundreds of times greater than comparable buried links. The dominant source of noise was shown to be caused by mechanical strains imposed by the swinging of the fiber spans due to transverse loading from the wind. Only the shortest link (32.6 km) was successfully stabilized for a period of nearly half an hour without a cycle slip, and with two orders of magnitude poorer stability than the equivalent buried link. Higher frequency division ratios in the servo error signal chain may improve the frequency locking performance of this and similar stabilization systems. In addition, repeater stations at intervals along the link will also improve the stability of the whole link. This increases the servo bandwidth and allows for greater servo gain. If the longer fiber spans can be successfully stabilized for significant periods of time, then aerial fiber may present a useful alternative for optical frequency transmission in situations where buried fiber is unavailable or cost-prohibitive. The challenges of robust signal stabilization over aerial fiber are significantly reduced at radio and microwave transfer frequencies due to the lower sensitivity of the phase of these transmissions to fiber length changes. The same aerial links tested here were successfully stabilized at radio [19] and microwave [20] frequencies.

**Funding.** University of Manchester, University of Western Australia (UWA).

**Acknowledgment.** We are very grateful to Jaco Müller, Roufurd Julie and Johan Burger for their efforts in supporting this field trip. Thank you to Charles Copley for providing the C-BASS weather station data. This paper describes work being carried out for the SKA Signal and Data Transport (SaDT) consortium as part of the Square Kilometre Array (SKA) project. The SKA project is an international effort to build the world's largest radio telescope, led by the SKA Organisation with the support of 10 member countries.